\documentclass[12pt]{article}
\usepackage{times}
\usepackage{geometry}
\usepackage[utf8]{inputenc}
\usepackage{enumitem,amssymb}
\usepackage{ragged2e}
\newlist{thematic}{itemize}{8}
\setlist[thematic]{label=$\square$}
\usepackage{pifont}

\usepackage{graphics,graphicx}

\usepackage{float}
\usepackage{lineno}
\usepackage{amssymb}
\usepackage[normalem]{ulem}

\begin{document}
\pagestyle{plain}
\pagenumbering{arabic}

\raggedright
\huge
Astro2020 Science White Paper \linebreak

Cosmic Rays and Interstellar Medium with Gamma-Ray Observations at MeV Energies \linebreak
\normalsize

\noindent \textbf{Thematic Areas:}  \hspace*{10pt} $\boxtimes$ Star and Planet Formation \hspace*{20pt} $\boxtimes$ Resolved stellar populations and their environments    \linebreak
   \linebreak
\textbf{Principal Authors:}
 \linebreak
 \linebreak
Name:	Elena Orlando
 \linebreak	
Institution: Kavli Institute for Particle Astrophysics and Cosmology and Hansen Experimental Physics Laboratory, Stanford University, CA (USA)
 \linebreak
Email: orlandele@gmail.com; eorlando@stanford.edu
 \linebreak
 \linebreak
 Name:	Isabelle Grenier
 \linebreak						
Institution: AIM, CEA, CNRS, Université Paris-Saclay, Université Paris Diderot, Sorbonne Paris Cité, F-91191 Gif-sur-Yvette, France
 \linebreak
Email: isabelle.grenier@cea.fr
 \linebreak
  \linebreak
 Name:	Vincent Tatischeff
 \linebreak						
Institution: CSNSM, CNRS/Univ. Paris-Sud, Universit\'e Paris-Saclay, Orsay, France
 \linebreak
Email: Vincent.Tatischeff@csnsm.in2p3.fr
 \linebreak
  \linebreak
 Name:	Andrei Bykov
 \linebreak						
Institution: Ioffe Institute, St.Petersburg, Russia
 \linebreak
Email: byk.astro@mail.ioffe.ru
 \linebreak
 
\textbf{Co-authors:} \\

Regina Caputo - NASA GSFC \\
Alessandro De Angelis - INFN and INAF Padova, Italy \\
Jurgen Kiener - CSNSM, CNRS/Univ. Paris-Sud, Universit\'e Paris-Saclay, Orsay, France \\
Alexandre Marcowith - Université de Montpellier \\
Julie McEnery - NASA GSFC \\
Andrew Strong - Max-Planck-Institut für extraterrestrische Physik \\
Luigi Tibaldo - IRAP, Universit\'e de Toulouse, CNRS, UPS, CNES, Toulouse, France \\
Zorawar Wadiasingh - NASA Goddard Space Flight Center \\
Andreas Zoglauer - University of California at Berkeley \\
\bigbreak

\textbf{Edorsers:} \\
Markus Ackermann - DESY \\
Andrea Addazi - Fudan University \\
Marco Ajello - Clemson University \\
Solen Balman - Self-Employed, Istanbul (at the moment, Earlier, METU, Ankara, Turkey) \\
Eugenio Bottacini - Universita' di Padova \\
Carl Budtz-Jørgensen - DTU Space, Technical University of Denmark \\
Sylvain Chaty - AIM, CEA, CNRS, Université Paris-Saclay, Université Paris Diderot \\
Rémi Chipaux - CEA DRF/IRFU \\
Paolo Coppi - Yale University \\
Flavio D'Amico - Instituto Nacional de Peqsuisas Espaciais - INPE (Brazil) \\
Filippo D'Ammando - INAF-IRA Bologna \\
Michaël De Becker - University of Liège \\
Roland Diehl - MPE Garching \\
Nicolao Fornengo - University of Torino and INFN/Torino \\
Luigi Foschini - INAF Brera Astronomical Observatory \\
Chris L. Fryer - Los Alamos National Laboratory \\
Daniele Gaggero - Instituto de Física Teórica UAM-CSIC \\
Nicola Giglietto - Politecnico di Bari and INFN-BARI \\
Nectaria Gizani - Helenic Open University, School of Science and Technology \\
Sylvain Guiriec - GWU/NASA GSFC \\
Elizabeth Hays - NASA GSFC \\
Kenji Hamaguchi - NASA/GSFC and UMBC \\
Andreas Haungs - Karlsruhe Institute of Technology, Germany \\
Andi Hektor - NICPB \\
John Hewitt - University of North Florida \\
Stefan Lalkovski - Sofia University "St. Kl. Ohridski" \\
Olivier Limousin - AIM, CEA, CNRS, Université Paris-Saclay, Université Paris Diderot \\
Manuela Mallamaci - INFN Padova \\
Antonino Marcianò - Fudan University \\
Philipp Mertsch - Institute for Theoretical Physics and Cosmology (TTK), RWTH Aachen University \\
Aldo Morselli - INFN Roma Tor Vergata \\
Josep M. Paredes - ICCUB, UB-IEEC, University of Barcelona \\
Asaf Pe'er - Bar Ilan University \\
Martin Pohl - University of Potsdam \\
Chanda Prescod-Weinstein - University of New Hampshire \\
Stefano Profumo - University of California, Santa Cruz \\
Luis Roso - Spanish Center for Pulsed Lasers, CLPU \\
Thomas Siegert - UCSD \\
Tonia Venters - NASA Goddard Space Flight Center \\
W. Thomas Vestrand - Los Alamos National Laboratory \\
Andrea Vittino - TTK, RWTH Aachen University \\

\justify

\textbf{Abstract: Latest precise cosmic-ray (CR) measurements and present gamma-ray observations have started challenging our understanding of CR transport and interaction in the Galaxy. Moreover, 
because the density of CRs is similar to the density of the magnetic field, gas, and starlight in the interstellar medium (ISM), CRs are expected to affect the ISM dynamics, including the physical and chemical processes that determine transport and star formation. 
In this context, observations of gamma-ray emission at MeV energies produced by the low-energy CRs are very important and urgent.\\
A telescope covering the energy range between $\sim$0.1 MeV and a few GeV with a sensitivity more than an order of magnitude better than previous instruments would allow for the first time to study in detail the low-energy CRs, providing information on their sources, their spectra throughout the Galaxy, their abundances, transport properties, and their role on the evolution of the Galaxy and star formation. Here we discuss the scientific prospects for studies of CRs, ISM (gas, interstellar photons, and magnetic fields) and associated gamma-ray emissions with such an instrument.}

\pagebreak



\section{Importance of Cosmic-Ray (CR) and Interstellar Medium (ISM) Studies} 
\vspace{-0.2cm} 

The Milky Way is very bright at gamma-ray energies. This is mainly due to the interactions of Galactic CRs with gas and photons producing hadronic pion-decay emission, and producing leptonic inverse-Compton scattering and bremsstrahlung emission, as CRs propagate from their sources throughout the Galaxy.
Observations of this gamma-ray emission from the Milky Way provide insights on the CRs spectra, density, distribution, transport and interactions properties, and the CR interplay with the ISM, even at large distances in the Galaxy (for extensive reviews see e.g. \cite{Strong2007,Grenier,Drury}).
Hence, observations from a few hundreds of KeV to a few tens of GeV allow us to investigate the properties of CRs and the ISM.  
However, observations so far by INTEGRAL, {\it Fermi} LAT, and COMPTEL underlined some discrepancies with present models, leaving open questions on the large-scale distribution of CR sources, on CR transport mechanisms in the Galaxy, and on their density and spectral variation over the Galaxy (see e.g. \cite{Grenier} and reference therein). Moreover, low-energy CRs are thought to be a fundamental component of the ISM, but their composition and flux are poorly known. In addition, the connection between low-energy CRs below a few GeV/nuc and galaxy evolution has started to be investigated only recently and is poorly understood. 
The energy density of Galactic CRs is similar to the energy density of interstellar gas, Galactic magnetic fields, and starlight. Hence CRs are an important component of the ISM. They impact the dynamics of the ISM, generating winds, and affecting physical and chemical processes that are responsible for the formation of stellar structures and galaxies (e.g. \cite{Ensslin, Persic}).
The ISM is very dynamic and its continuous transformations in phase and in density control the efficiency by which galaxies consume their gas to form stars. Sub-GeV CRs play a significant role in this evolution by ionizing the gas, e.g.  \cite{Padovani}, thereby heating the darkest and densest clouds, as well as initiating chemical reactions that include the production of gas coolants \cite{Grenier}. 
GeV CRs, on the other hand, provide pressure support in rough equipartition with the thermal pressure \cite{Pfrommer2017}, as observed locally, and CR pressure gradients help push the gas blown by supernovae out of Galactic discs \cite{Pakmor16}.


In general, there is a common interest in understanding CRs at energies below $\sim$100~GeV/nuc. This paper presents the scientific topics and expected outcomes that a mission from a few hundreds of keV to a few tens of GeV can target with the goal of understanding the role of CRs in the galaxies. A mission at MeV energy range would allow for the first time to study in detail the CRs with energies below a few GeV/nuc, which play a fundamental role in the formation of stars and on the dynamics in the Galaxy. CR sources, acceleration mechanisms, injection spectra into the ISM, transport properties, and their spectral and spatial distribution in the entire Galaxy, and in specific regions such as molecular clouds, star forming regions, and stellar clusters would be investigated for the first time for the entire energy band, and with unprecedented sensitivity and spatial resolution. As a consequence, distribution, acceleration, transport, and effects of CRs on the ISM and on the dynamics and evolution of the Galaxy can be finally understood. \\
{\bf This is of interest of many proposed missions at MeV energies, including AMEGO \cite{amego}, e-ASTROGAM \cite{eastrogam} (and the All-Sky ASTROGAM), COSI \cite{cosi}, and GalCenEx \cite{galcenex}.}

\vspace{-0.3cm}
\section{Specific topics in more detail}
\vspace{-0.2cm}

In this section we describe each scientific topic that would be addressed with such a telescope.  \\

\vspace{-0.1cm}
\subsection{Large-scale interstellar emissions}
\vspace{-0.1cm}


\begin{figure}[h!]
\centering
\includegraphics[width=0.48\textwidth]{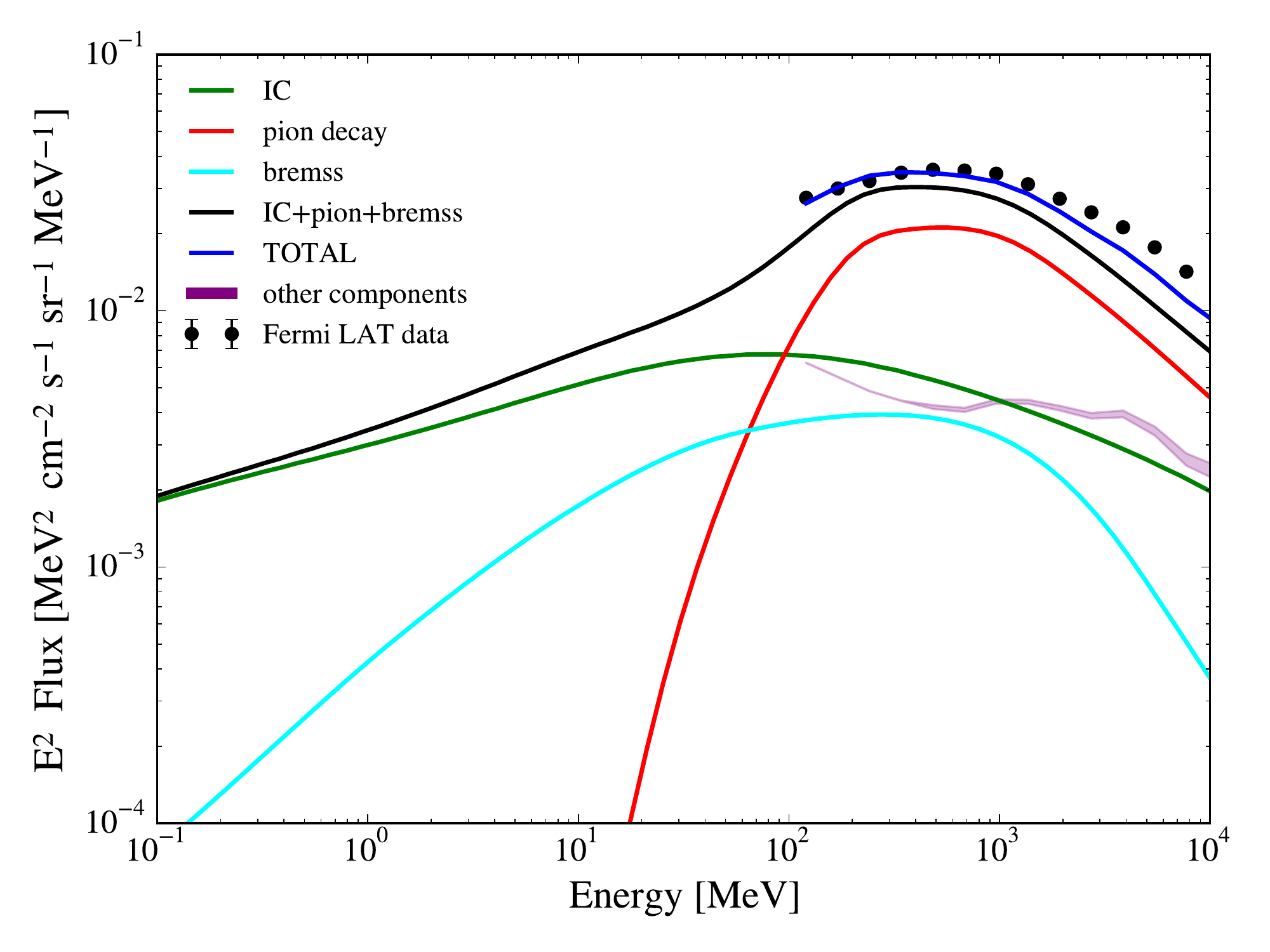}
\includegraphics[width=0.48\textwidth]{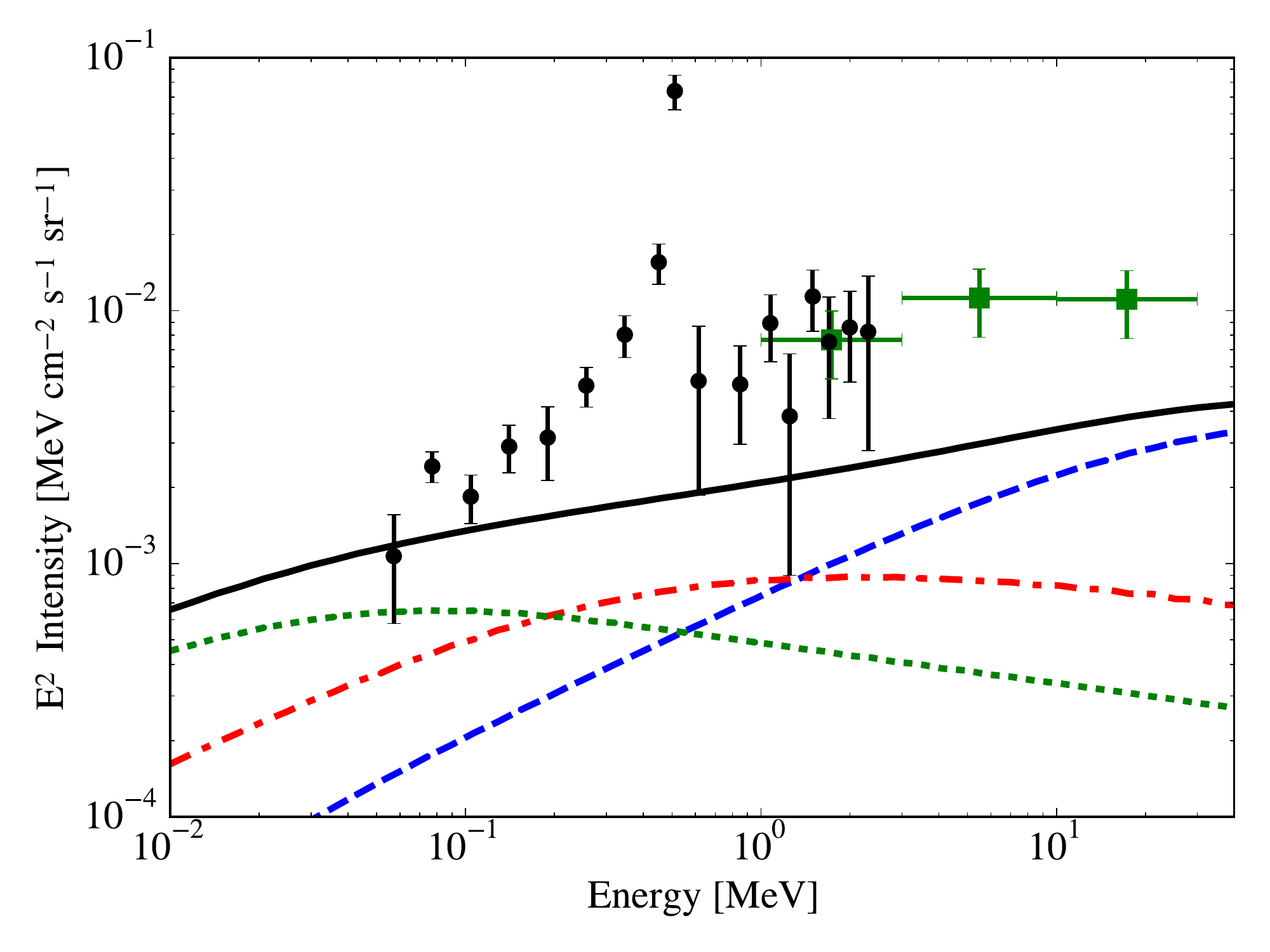}
\vspace{-0.7cm}
\caption{\label{fig:fig1}Predictions of the large-scale interstellar emission from 10 KeV to 10 GeV for the best model of \cite{O2018} that fits latest Voyager 1 AMS02 data, and synchrotron radio-microwave data. Figure on the left, adapted from \cite{O2018}, shows the IC (green line), bremsstrahlung (cyan line), and pion decay (red line) model, along with their sum
(black line) compared with the {\it Fermi} LAT data for a 10$^o$ radius around the Galactic center (black points), while the figure on the right, as in \cite{O2018}, shows the predictions for the IC on the CMB (green dotted line), on the diffuse IR (red, dash-dotted line), and on the diffuse optical (blue dashed line), along with their sum
(black solid line) compared to the INTEGRAL/SPI (black points) and COMPTEL (green points) data for the region $|b|<$15$^{\circ}$ and $|l|<$30$^{\circ}$.  An AMEGO-like instrument sensitivity is below the scale of the plotted area. {\it Fermi} LAT data are from \cite{FermiGC}, SPI data from \cite{Bouchet2011}, COMPTEL data from \cite{Strong99}.}
\end{figure}

Among important discoveries enabled by COMPTEL, INTEGRAL/SPI, and {\it Fermi} LAT, we are facing the difficulty of disentangling the different components (leptonic and hadronic) of the interstellar gamma-ray emission produced by CRs, and even on model degeneracies between truly diffuse emission and emission by unresolved sources. As a consequence deriving further details on CRs and their transport in the ISM is still challenging, if not impossible, with present telescopes. 
For example, {\it Fermi} LAT data have shown that the gamma-ray emission decreases in flux \cite{diffuse2} and softens \cite{Fermi2016} from the inner Galaxy to large Galactocentric radii differently to what is expected from the distribution of potential CR sources and under the assumption of uniform diffusion. A large halo size, additional gas and CR sources in the outer Galaxy \cite{diffuse2}, anisotropic diffusion coefficient (e.g. \cite{DeMarco2007,Tomassetti,Pakmor2016a,Cerri2017}), advection (e.g. \cite{Recchia2016, Pakmor16, Evoli}, all might provide possible solutions. Indeed, there are further limitations to our knowledge of CRs, such as the distribution of CRs in the Galaxy (e.g. if they are more concentrated in the spiral arms or in the Galactic center), or how they are affected by Galactic winds \cite{Pfrommer2017} and by the still uncertain Galactic magnetic field \cite{Hanasz2009,Giacinti2012,Girichidis,Orlando2019}. Indeed CRs can also drive winds (e.g. \cite{Breitschwerdt1991, Recchia2017} that, in turn, affect the CR large-scale distribution. 
Moreover, the usual assumption that direct CR measurements, after accounting for solar modulation, are representative of the proton local interstellar spectrum in our $\sim$1~kpc region is also challenged \cite{O2018}.
Among the open questions from {\it Fermi} LAT data, the discovery of the Fermi Bubbles is still puzzling scientists (e.g. \cite{Su,FermiBubbles}). If they are due to  past activity of a supermassive black hole at the center of the Milky Way (AGN scenario) or due to a period of starburst activity (starburst  scenario) is unclear (see \cite{eastrogam} and references therein for a detailed review). Moreover, smaller structures such as excesses and dips are present in the {\it Fermi} LAT data \cite{FermiGC} after subtracting the models, possibly due to some gas emission, unresolved sources, or sites of CR acceleration. 
{\it Fermi} LAT has also detected gamma-ray emission from nearby galaxies \cite{FermiGalaxies}, where CRs have just started to be investigated. 

{\bf Importance and expected results} \\
The  hadronic gas-related pion-decay emission is the major interstellar component at GeV, while below 100 MeV most of the emission comes from IC scattering and  from bremsstrahlung  emission  due  to  CR  electrons. Indeed, IC is predicted to be the dominant interstellar component below a few tens of MeV \cite{O2018, Bouchet2011, Strong2011}. Figure~1 shows the expected spectra of the different components for a model \cite{O2018} that fits latest Voyager 1 and AMS02 data, and also synchrotron interstellar emission as observed in radio and microwave. 
A telescope at MeV energies, such as AMEGO or All-Sky ASTROGAM, will for the first time allow to clearly discriminate emission from CR nuclei and electrons. 
Moreover, it will also reveal the spatial and  spectral distributions  of  the IC  emission  in  the Galaxy, important for disentangling emission from unresolved sources, the extragalactic diffuse gamma-ray background,
or potential dark-matter signal. This also allow to infer the distribution of CR electrons, which better sample CR inhomogeneity, because they are affected by energy losses more strongly than nuclei, and they remain much closer
to their sources. Moreover, observations of gamma rays below 100 MeV by the same electrons that produce synchrotron emission in radio and microwaves provide firmer constraints on Galactic magnetic fields (e.g. \cite{Strong2011b,O2013,Planck,Imagine}).


\vspace{-0.3cm}
\subsection{ISM in dense regions}
\vspace{-0.2cm}

Due to the limited resolution and energy coverage of past and present instruments we lack crucial observational constraints on the inhomogeneities of the low-energy CR distribution in and around star-forming regions, on their penetration into dense clouds \cite{Dogiel18, Gabici}, and on their potential production by protostar jets within cloud cores \cite{Dickinson,Padovani16}.
Conversely, we don't know how much of an imprint star-forming regions leave on the CR distributions at all energies. We expect this imprint to be significant, first because CR sources are clustered in space and time around their parent OB associations, second because OB associations may impart a fraction of the kinetic energy of their supersonic stellar winds to CR acceleration or re-acceleration \cite{Bykov01}, and third because stellar-induced MHD turbulence should impact CR diffusion \cite{Lazarian12}.

Our views on the diffusion properties of Galactic CRs have largely been inferred within 1 or 2 kpc, and at energies larger than several GeV which correspond to diffusion speed of hundreds of parsecs per Myr. How biased is our viewpoint from within the mini-starburst Gould Belt? How do low-energy CRs propagate in and out of such regions? These are central questions to be answered primarily in low-energy $\gamma$ rays in order to better understand the CR feedback on the multi-phase structure of the ISM. 

\begin{figure}[h!]
\centering
\includegraphics[width=0.9\textwidth]{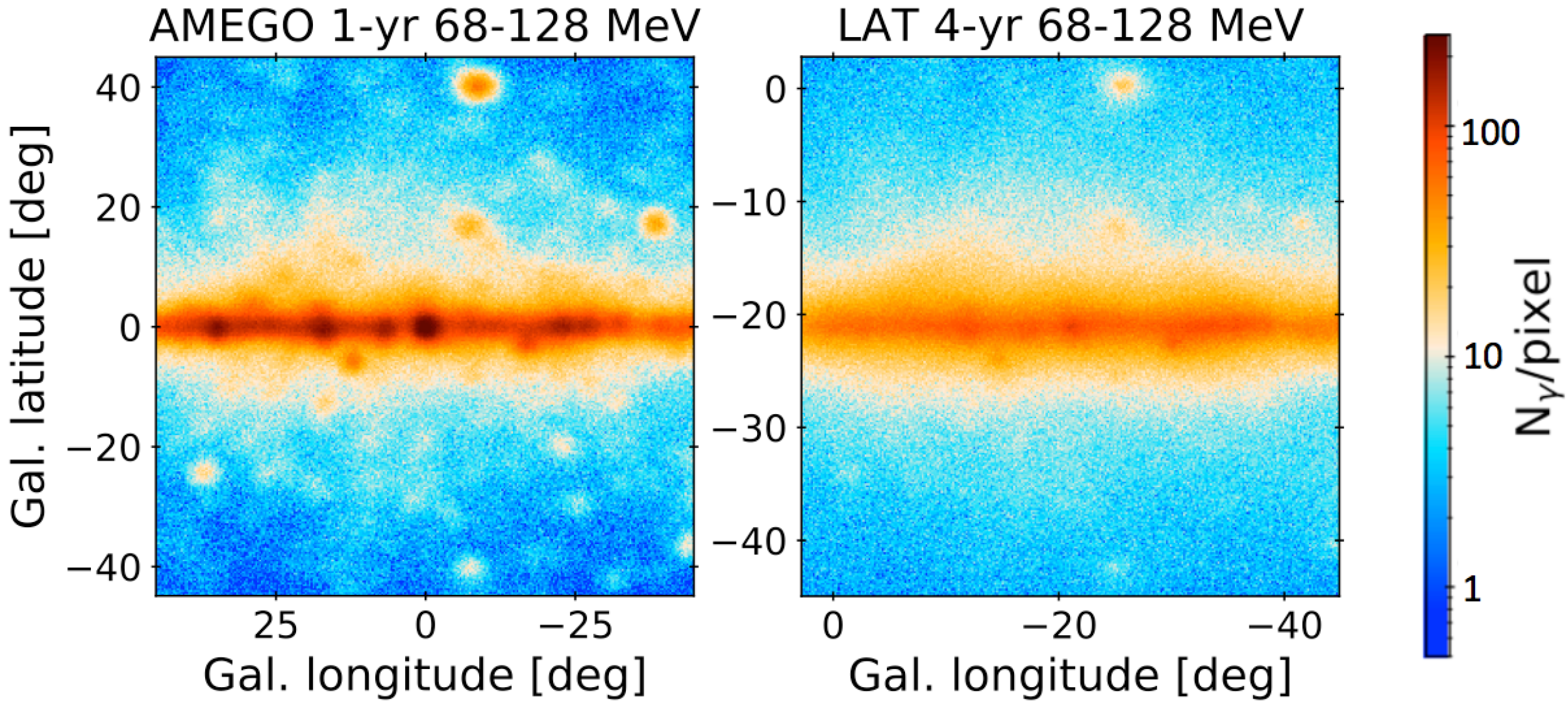}
\vspace{-0.5cm}
\caption{\label{fig:fig2}Photon maps  
of the inner Galactic regions in 68 to 127 MeV energy band
simulated for 1 year of effective exposure of AMEGO (left) and observed for 4 years by {\it Fermi} LAT (right). The AMEGO simulations are based on the {\it Fermi} LAT interstellar model, including the Fermi Bubbles, and on the spectra of the 4FGL $\gamma$-ray point sources.}
\end{figure}

{\bf Importance and expected results} \\
These issues need to be addressed by spatially resolving the MeV to TeV emission produced in and around a variety of star-forming regions, powered by stellar clusters of different masses and at different stages of evolution (e.g. \cite{Gaches}. An instrument such as AMEGO will be pivotal in synergy with the HAWC and CTA observatories at TeV energies, with the {\it Fermi} LAT archive, and with MeerKAT and SKA. Severe confusion has prevented low-energy CR studies so far, even with {\it Fermi} LAT. Figure~2 shows that the improved performance of an AMEGO-like instrument will open new avenues to (i)  probe CRs in nearby clouds off the Galactic plane down to a half parsec scale above 1 GeV and 10 pc at 50 MeV, (ii) compare the CR proton and electron spectra in and around nearby OB associations (e.g. in the Orion and Rosette nebulae), and (iii) to search for enhanced $\gamma$-ray activity in massive OB associations beyond the few cases discovered at higher $\gamma$-ray energies, such as the single robust  case of Cygnus X \cite{FermiCygX}.  
The performance of an AMEGO-like instrument will be key to reliably extend the spectra of the CR-induced interstellar emissions below 500 MeV in order to measure the energy distribution of the bulk of the CR nuclei, to estimate the CR pressure inside OB superbubbles \cite{FermiCygX}, to follow the release and diffusion of CR electrons and nuclei around supernova remnants (e.g. \cite{Nava,Vaupre}, and to measure how low-energy CRs get depleted inside dense clouds because of self-excited MHD turbulence \cite{Dogiel18}.
An AMEGO-like instrument will also search for $\gamma$-ray counterparts to synchrotron emitting protostars to set limits on CR production by jets which impacts the ionisation, therefore the evolution of protostellar discs \cite{Padovani16}.

\vspace{-0.3cm}
\subsection{De-excitation lines} 
\vspace{-0.1cm}
The {\it Voyager 1} spacecraft has recently provided valuable measurements of the local energy spectra of Galactic CR nuclei down to $\sim 3$~MeV~nucleon$^{-1}$  beyond the heliopause \cite{cum16}, but the total CR ionization rate of atomic hydrogen resulting from the measured spectra is a factor $>10$ lower than the average CR ionization rate measured in clouds across the Galactic disc using line observations of ionized molecules by Herschel \cite{neu17}. The difference suggests that low energy CRs are relatively less abundant in the local ISM than elsewhere in the Galaxy. Observations of  H$_3^+$  in diffuse molecular clouds show indeed that the density of low energy CRs can strongly vary from one region to another in the Galactic disk, and, in particular, that the low energy CR flux can be significantly higher than the average value in diffuse molecular gas residing near a site of CR acceleration such as a supernova remnant \cite{ind10,ind12}. 

A very promising way to study the flux and composition of CR nuclei below the kinetic energy threshold for production of neutral pions ($\approx 300$~MeV for $p+p$ collisions) would be to detect gamma-ray lines in the $0.1 - 10$~MeV range produced by nuclear collisions of CRs with interstellar matter. Strong narrow lines are produced by the excitation of abundant heavy nuclei of the ISM by CR protons and $\alpha$-particles with kinetic energies between a few MeV and a few hundred MeV. The most intense lines are expected to be the same as those frequently observed from strong solar flares, i.e. lines from the de-excitation of the first nuclear levels in $^{12}$C, $^{16}$O, $^{20}$Ne, $^{24}$Mg, $^{28}$Si, and $^{56}$Fe \cite{ram79}. The total nuclear line emission is also composed of broad lines produced by interaction of CR heavy ions with ambient H and He, and of thousands of weaker lines that together form a quasi-continuum in the range $E_\gamma \sim 0.1 - 10$~MeV \cite{ben13}. Some of the prominent narrow lines, e.g. that at 6.13~MeV from $^{16}$O, may exhibit a very narrow component from interactions with dust grains, where the recoiling excited nucleus can be stopped before the gamma-ray emission \cite{tat04}. 

\begin{figure}
\centering
\includegraphics[width=0.45\textwidth]{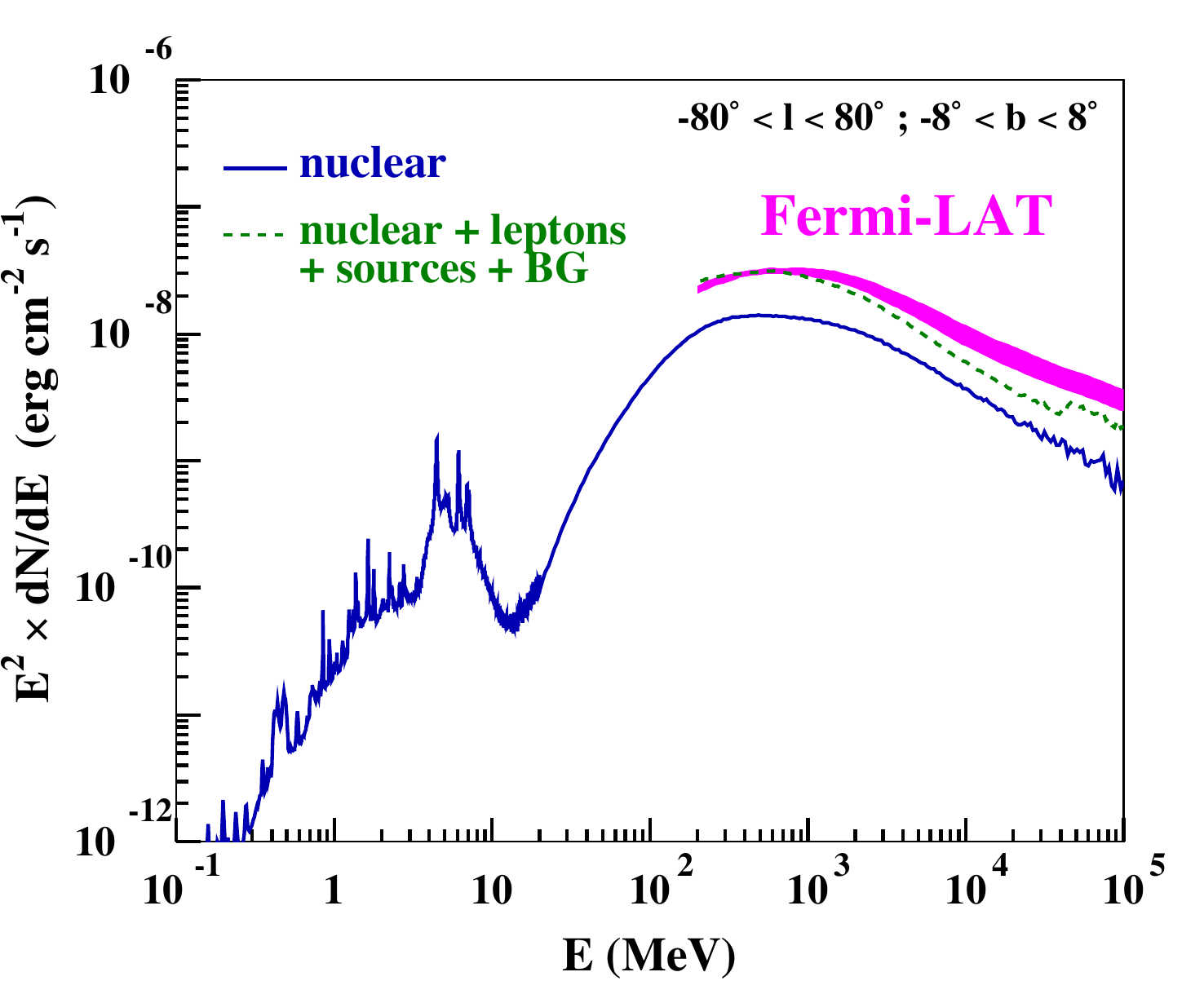}
\includegraphics[width=0.44\textwidth,scale=1.2]{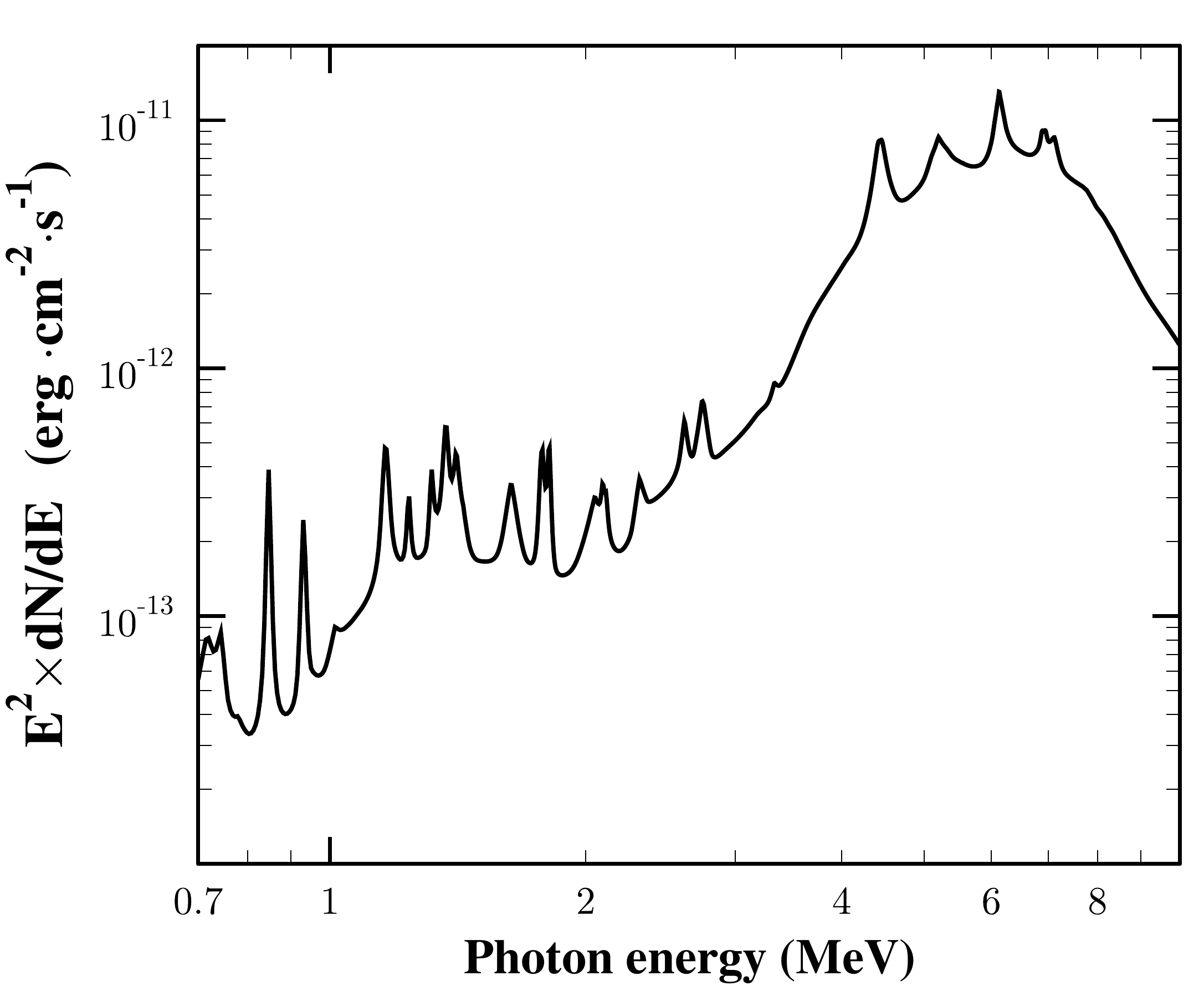}
\vspace{-0.5cm}
\caption{Left panel: Predicted gamma-ray emission due to nuclear interactions of CRs in the inner Galaxy (longitude $-80^\circ \leq l \leq 80^\circ$ and latitude $-8^\circ \leq b \leq 8^\circ$). The gamma-ray line emission below 10 MeV is due to low energy CRs, whose properties in the ISM have been adjusted such that the mean CR ionization rate deduced from H$_3^+$ observations and the  \textit{Fermi}-LAT data (magenta band) at 1 GeV are simultaneously reproduced (adapted from \cite{ben13}). The dashed green line shows the total calculated emission when adding leptonic contributions, point sources and extragalactic gamma-ray background that were taken from \cite{diffuse2}. Right panel: A simulated spectrum of gamma-ray line emission from a Galactic superbubble of a scale size about 50 pc located at a distance about 500 pc.  A few Myrs age superbubble is powered by multiple OB star winds and supernovae accelerating CRs \cite{Bykov01}.}
\label{fig:fig4}
\vspace{0pt}
\end{figure}

{\bf Importance and expected results} \\
Figure~\ref{fig:fig4} (left panel) shows the calculated gamma-ray emission spectrum from CRs in the inner Galaxy containing a low-energy component that would account for the observed mean ionization rate of diffuse molecular clouds. Observations of this emission would be the clearest proof of an important low-energy CR component in the Galaxy and probably the only possible means to determine its composition, spectral and spatial distribution. A particularly promising feature of the predicted gamma-ray spectrum is the characteristic bump in the range $E_\gamma = 3 - 10$~MeV, which is produced by several strong lines of $^{12}$C and $^{16}$O. The calculated flux in this band integrated over the inner Galaxy ($\mid$$l$$\mid \leq 80^\circ$; $\mid$$b$$\mid \leq 8^\circ$) amounts to $7 \times 10^{-5}$~photons~cm$^{-2}$~s$^{-1}$. In comparison, in three years of nominal mission lifetime, the predicted sensitivity of an AMEGO-like instrument for such a spatially extended emission is $S_{3\sigma}\sim 5 \times 10^{-6}$~photons~cm$^{-2}$~s$^{-1}$. In Fig.~\ref{fig:fig4} (right panel) a simulated gamma-ray line spectrum  of an individual nearby superbubble is shown \cite{Bykov01}. The spectrum is dominated by both narrow and broad $^{12}$C and $^{16}$O lines providing a way to constrain low energy CR composition. 


\pagebreak

\def \aap  {A\&A}
\def \aaps  {A\&AS}
\def \aj  {AJ}
\def \apj  {ApJ}
\def \apjs  {ApJS}
\def \apjl  {ApJL}
\def \aplett  {ApJL}
\def \apss  {AP\&SS}
\def \araa  {ARA\&A}
\def \jcap  {JCAP}
\def \prd {Phy. Rev. D}
\def \ssr {SSRv}
\def \mnras {MNRAS}
\def \nat {Nature}
\def \physrep {Phys. Rept.}
\def \pasj {PASJ}
\def \etal {et~al.~}
\def \rmxaa {RMxAA}
\def \jgr {JGR}
\def \pasp {PASP}


\end{document}